\documentclass[pre, twocolumn, superscriptaddress, showpacs, floatfix]{revtex4}
\usepackage{amsmath}
\usepackage{amsfonts}
\usepackage{amssymb}
\usepackage{graphicx}
\usepackage{epstopdf}
\usepackage{multirow}

\begin{document}

\title{Avalanche localization and crossover scaling in amorphous plasticity}

\author{Zoe Budrikis}
\affiliation{ISI Foundation, Via Alassio 11/c, 10126 Torino, Italy}
\author{Stefano Zapperi}
\affiliation{CNR-IENI, Via R. Cozzi 53, 20125 Milano, Italy}
\affiliation{ISI Foundation, Via Alassio 11/c, 10126 Torino, Italy}

\begin{abstract}
We perform large scale simulations of a two dimensional lattice model for amorphous plasticity with random local yield stresses and long-range quadrupolar elastic interactions. We show that as the external stress increases towards the yielding phase transition, the scaling behavior of the avalanches crosses over from mean-field theory to a different universality class. This behavior is associated with strain localization, which significantly depends on the short-range properties of the interaction kernel. 
\end{abstract}


\maketitle
\section{Introduction}
In recent years, strong experimental evidence has emerged that the plastic response of materials is not smooth and continuous --- as would be expected from classical stress-strain relationships --- but instead intrinsically subject to fluctuations \cite{Zaiser2006}, both temporally and spatially. For example, upon application of external stress, plastic strain in micron sized samples increases in avalanches with power law distributions~\cite{Dimiduk2006}. Such avalanches are also present in bulk materials, as shown by acoustic emission measurements~\cite{miguel2001}. Furthermore, acoustic emission localization \cite{weiss2003} and surface observations \cite{Zaiser2004} reveal complex spatial patterns.

In crystalline materials, the observed intermittent behavior is related to the dynamics of interacting dislocations \cite{miguel2001,csikor2007}, and it is believed that because interactions between dislocations are long-ranged, the scaling of this intermittent behavior is of a mean-field nature \cite{Zaiser2005,Zaiser2007,Dahmen2009,Dahmen2011}. 
At the same time, experiments \cite{shan2008,Wang2009,Sun2010,Sun2011,Sun2012} and molecular dynamics simulations \cite{maloney2004,demkowicz2005,maloney2006,Bailey2007,Lemaitre2009,Maloney2009, Salerno2012} show that scale free avalanches and complex strain patterns are also shared by amorphous materials where 
localized defects are not present. An open question that remains is whether the universality class of amorphous plasticity is the same as that of crystalline plasticity.

In two dimensions, amorphous plasticity can be captured by a simple model 
\cite{Baret2002,Picard2005,Talamali2011,Talamali2012,Vandembroucq2011} with two competing ingredients:  (i) disorder in the form of randomly distributed local yield thresholds, and (ii) a long-range, anisotropic interaction kernel, which in an infinite system is given by the Eshelby form $\cos(4\theta)/r^2$
\cite{Eshelby1957}.  The build up of plastic strain can be mapped on to the motion of a 2D interface in the transverse direction, driven by external stress. The random yield thresholds represent a landscape of random energy barriers through which the interface moves, so that the model can be thought of as a type of $2+1$ dimensional depinning model that undergoes a transition from the pinned phase to the moving phase as external stress is increased \cite{Baret2002,Picard2005,Talamali2011,Talamali2012,Vandembroucq2011}. 

In fact, a formally similar model was derived in Ref. \cite{Zaiser2005} as a mesoscale model for crystal plasticity.
The similarity derives from the fact that in two dimensions the stress created by moving a dislocation
in a crystal is equivalent to the one produced by a localized shear event in an amorphous material.
Therefore we could expect that crystal and amorphous plasticity share the same critical behavior,
at least in two dimensions, where mean field behavior has been observed \cite{Zaiser2005}.
 This is, however, contradicted by recent results by Talamali {\it et al}~\cite{Talamali2011} suggesting that at criticality, the amorphous plasticity model is not mean-field, as evidenced by, e.g, avalanche size distributions.

To shed some light on the universality class of amorphous plasticity, we present large-scale simulation studies of plasticity in 2D amorphous materials. Previous studies of this system have suffered from small system sizes, a consequence of the large computational cost induced by the long-range interactions. By implementing our simulations on the parallel architecture of a graphics processing unit, we are able to access larger system sizes than those previously published, enabling a clearer picture to be built up of the critical behaviour of the model. 

This paper is organized as follows. First, we examine the strain avalanches of the system away from criticality, and find that --- depending on the short-range part of the interaction kernel --- the model can exhibit a nonuniversal crossover from mean-field behavior at small stresses to non mean-field behavior, similar to that observed in a lattice model for ductile fracture~\cite{picallo2010}. We then examine the non mean-field behavior at criticality in greater depth, and present evidence that it is indeed universal, depending only on the large-scale nature of interactions. The onset of non mean-field behavior is connected to strain localization, which we find is affected by small changes to short-range interactions. This is especially apparent when weakening is introduced to the model.

\section{Model}
\label{methods}
We consider a two-dimensional square lattice, in which each lattice site represents an area element small enough that the strain is uniform within it. Unless otherwise stated, all results presented here are for systems of size $L=1024$ with periodic boundary conditions. This is a matter of some subtlety, because it is essential to ensure that the mean of the interaction kernel is zero over the whole system, to avoid imposing an extraneous net stress.
Here, we consider two periodization methods, which are distinguished by their short-range behaviors. As discussed below, the nonuniversal effects caused by these differences can be considerable.

\begin{figure}
\includegraphics[width=\columnwidth]{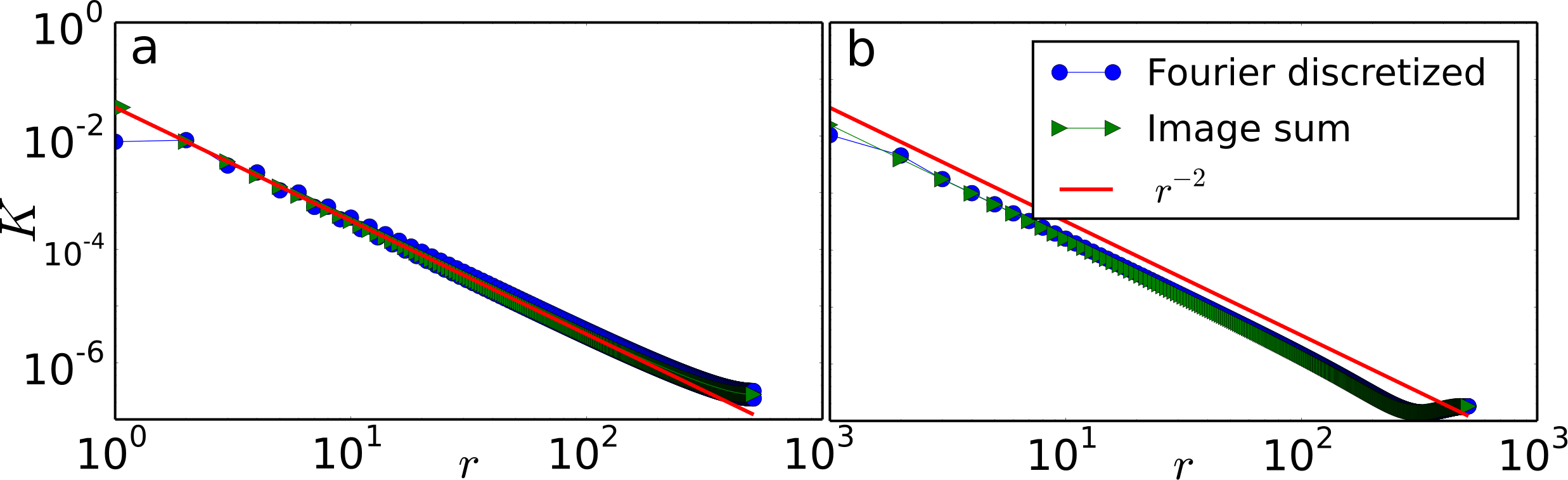} 
\caption{\label{kernels} Magnitude of the interaction strength, $K$, between pairs of sites separated in real space by distance $r$, along lines in the (a) $x$ and (b) $y=x$ direction. The two methods of periodization give almost identical results, except for short-range interactions. The red lines indicate a $1/r^2$ decay.}
\end{figure}

In the first method, an infinite set of images of the system in the $y$ direction are summed over analytically, and then a finite number of exponentially-decaying terms arising from images in the $x$ direction are summed numerically, as described in Appendix~\ref{ImageSum}. We refer to this periodized kernel as the ``image sum kernel''. We emphasize that the summation is a purely formal procedure for obtaining an interaction kernel that is periodic with the same short-range behaviour as the kernel in the infinite system, since the sum over images is in fact only conditionally convergent~\cite{Cai2003}. This method conserves the $1/r^2$ nature of the original (infinite system) kernel at small distances, as shown in Fig.~\ref{kernels}. 

The second method of periodization, which has been studied previously~\cite{Talamali2011, Talamali2012}, is to Fourier transform the kernel analytically in the infinite system, and then discretize this on an $L\times L$ lattice in Fourier space, setting the $k=0$ term (that is, the mean of the kernel in real space) to zero. Transforming back to real space using a discrete Fourier transform ensures periodization. We refer to this kernel as the ``Fourier kernel''. We emphasize that this method of periodization, like the image summation method, introduces discretization effects, because high-frequency modes are cut off. A side-effect of this is to modify the short-range part of the interaction away from a $1/r^2$ form, as illustrated in Fig.~\ref{kernels}. However, both kernels are nearly identical at large distances, following $1/r^2$ except near the system edges where periodization imposes a distortion; and both kernels are also stable, with all Fourier modes non-positive.

The system is initialized with strain $\gamma_i=0$ at all sites, and stress thresholds $t_i$ taken from a uniform distribution over $[0,1)$. The external stress $f$ is increased adiabatically, and in each ``time step'' all sites that reach their threshold are updated simultaneously, by increasing their strain by a fixed amount $d\gamma=0.1$ and taking a new yield threshold. Unless otherwise stated, yield thresholds are always chosen from the same distribution, that is, there is no hardening or softening. An advantage of this method of driving is that avalanches are well defined. The size of an avalanche is the total strain increase that occurs before the external stress is increased again, and the duration is the number of time steps taken. The simulation is stopped when the system reaches a mean strain of $\gamma_T$, which is large enough to avoid spurious size effects~\cite{Budrikis2013} (the numerical value of $\gamma_T$ is typically $100\%$--$150\%$). The parallelization of this algorithm  for implementation on a graphics processing unit is discussed in Appendix~\ref{gpu}.

\section{Results}

\subsection{Nonuniversal crossover from mean field behavior}
\begin{figure}
\includegraphics[width=0.9\columnwidth]{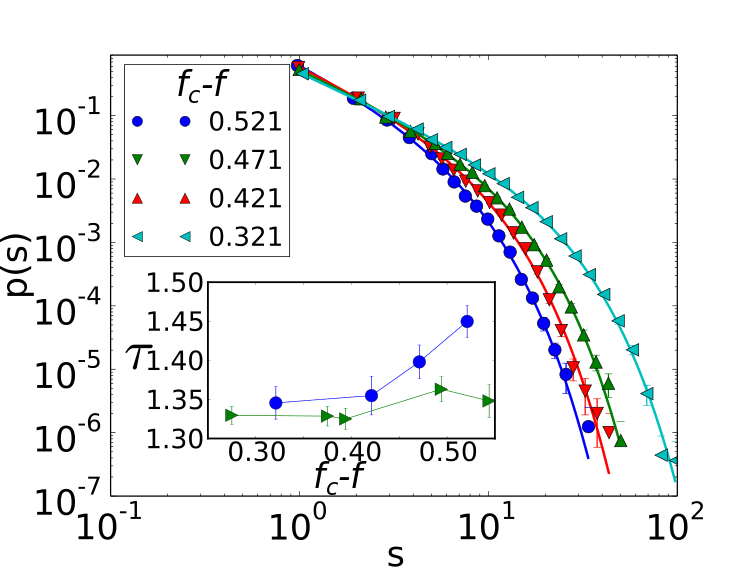} 
\caption{\label{fit_small_stress} Avalanche size distributions, $P(S)$ for a system governed by the image sum kernel. Distributions are measured for stress bins with width $\Delta f=0.01$ and mean values (relative to critical stress $f_c=0.626$) as indicated to the legend. The solid lines are fits, using the form given in Eq.~\eqref{LDW}. Inset: dependence of avalanche size exponent $\tau$ on stress for the image sum kernel (blue circles) and the Fourier discretized kernel (green triangles). As criticality is approached, the exponents become indistinguishable.}
\end{figure}

\begin{figure}
\includegraphics[width=\columnwidth]{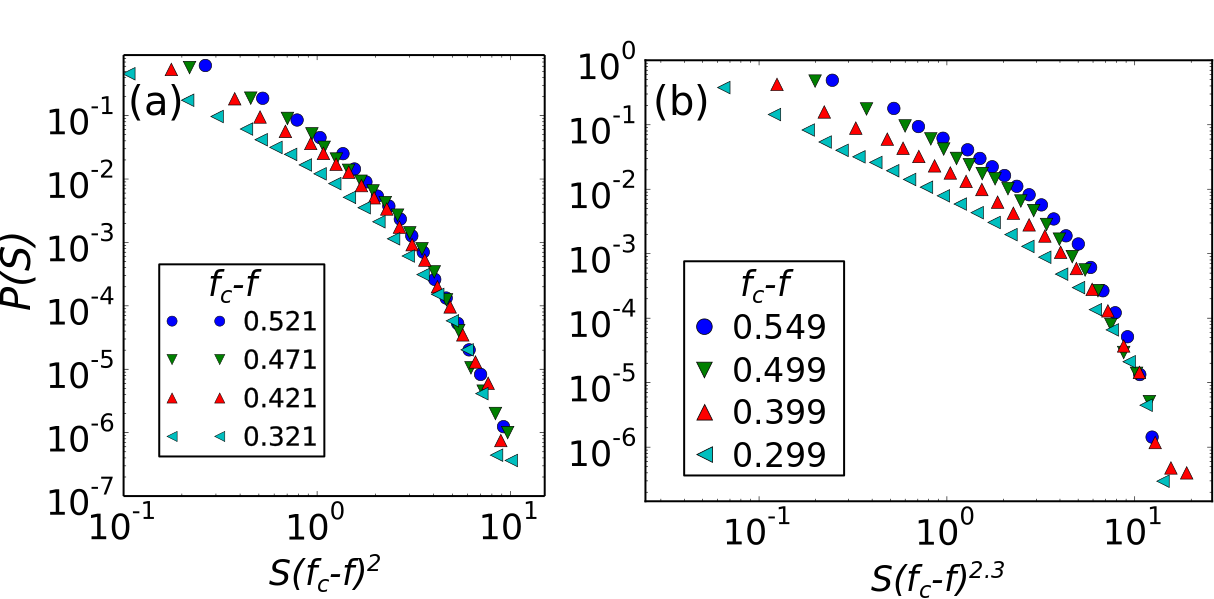} 
\caption{\label{tail_collapse_small_stress} (a)  Collapse of tails of the size distribution for the image sum kernel upon rescaling by $(f_c-f)^2$, in accordance with mean field theory. (b) For a system governed by the Fourier discretized kernel, the tails of the avalanche size distributions collapse upon rescaling by $(f_c-f)^{2.3}$}
\end{figure}

We first examine the strain avalanche behavior of the system at small applied stresses, away from criticality. In these simulations, the external stress is the control parameter and is held constant during each avalanche. The avalanche size distributions are measured for stress bins (of width $\Delta f=0.01$) and are shown in Fig.~\ref{fit_small_stress}. The distributions can be fit using a function based on that given by Le Doussal and Wiese~\cite{LeDoussal2012b} for a one-loop correction to mean-field theory depinning:
\begin{equation}
\label{LDW}
P(S) = c_1 S^{-\tau} \exp\bigl(c_2 (-B S^{-\delta} + C \sqrt{S}) \bigr)
\end{equation}
where $c_1$ and $c_2$ are fitting parameters that absorb the normalization, and $B$, $C$ and $\delta$ are related to $\tau$ via the parameter $a$: $\tau =3/2 + 3a/8$, $B = 1 - a (1+\gamma_E/4)$, $C = - \sqrt{\pi} a/2$, $\delta = 1 -a/4$, where $\gamma_E$ is Euler's constant. For a mean field model, $\tau=3/2$, $\delta=1$ and $C=0$. We emphasize that we have selected this model for purely empirical reasons --- it fits our data for small-stress simulations well and has a mean-field limit --- rather than because we expect it to necessarily be theoretically valid. 

Considering first the image sum kernel, we find that the exponent $\tau$ drifts from a measured value of $1.45\pm0.02$ for stresses in $[0.1,0.11]$ (to be compared to the critical stress $f_c=0.626$, as determined from finite size scaling), to $\tau\approx 1.35$ as criticality is approached. The dependence of $\tau$ on stress is shown in the Inset of Fig.~\ref{fit_small_stress}. 

The variation in fitted $\tau$ is not, in itself, proof of a crossover from mean-field to non mean-field behavior, since the avalanche distributions are rather narrow at small stresses and the change in $\tau$ could be spurious. However, further evidence for mean field behavior is given by the scaling of the cutoff of $P(S)$ with stress. The location of the cutoff, $S^*$, is dictated by how close the external stress is to its critical value, and is given by the scaling relation
\begin{equation}
S^* \propto |f_c-f|^{-1/\sigma}
\end{equation}
where $1/\sigma$ takes the value $2$ in the mean field universality class. Figure~\ref{tail_collapse_small_stress}(a) shows the collapse of the tails of $P(S)$ for small stresses when $S$ is rescaled in this way.

On the other hand, in contrast to the image sum kernel, at small stresses the Fourier discretized kernel does not give rise to a mean field regime. As shown in the Inset of Figure~\ref{fit_small_stress}(a), fitting the distributions $P(S)$ (obtained for stress bins of width $\Delta f=0.01$) with the form given in Eq.~\eqref{LDW} yields an exponent $\tau \sim 1.35$, with no clear stress dependence, even at small stresses. Furthermore, as shown in Fig.~\ref{tail_collapse_small_stress}(b), for small stresses the cutoffs of the avalanche size distributions scale as $(f_c-f)^{-2.3}$, which is also not consistent with mean field theory. It therefore seems likely that if mean field behavior does occur, it is only for extremely small stresses.

\subsection{Universal, non mean-field behavior at criticality}

To better understand this system at criticality, we perform strain-controlled simulations in which the plastic system is coupled to a spring of stiffness $k$.  The total strain in the system, $\gamma_{\rm tot}$ --- sum of the plastic strain $\gamma$ and the spring extension --- is increased adiabatically. In the absence of a plastic subsystem, $\gamma_{\rm tot}$ would be attained at an external stress $f=k\gamma_{\rm tot}$, following Hooke's law. However, the presence of the plastic system reduces the spring extension by $\gamma$, so that the external stress is instead $f = k (\gamma_{\rm tot} - \gamma)$.
In these simulations, the plastic system can be maintained indefinitely in a critical steady state, allowing us to collect detailed avalanche statistics. 

\begin{figure}
\centering
\includegraphics[width=\columnwidth]{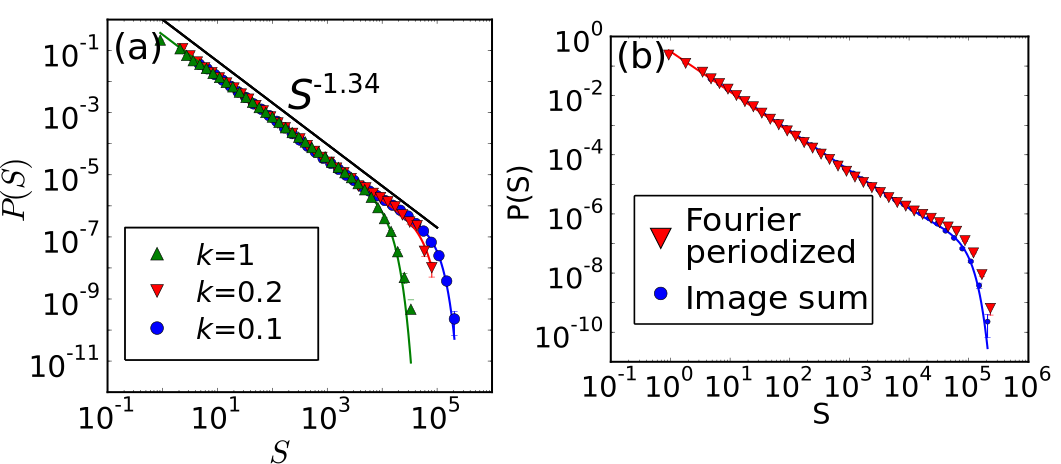} 
\caption{\label{avalanches_steady_state} Avalanche size distributions, measured in the steady state of strain controlled simulations. Part (a) shows results for the image sum kernel, for various spring constants $k$. Part (b) shows results for both kernels, for $k=0.1$. In both cases, the solid lines are fits obtained using the form given in Eq.~\eqref{steadystate}. }
\end{figure}

The avalanche size distributions in this case are decidedly not mean-field, and have a distinct ``bump'' in the large-avalanche tail. They cannot be fit using the form given by Le Doussal and Wiese, and instead we fit with
\begin{equation}
\label{steadystate}
P(S) = c_1 S^{-\tau} \exp(c_2 S - c_3 S^2),
\end{equation}
with $c_1$, $c_2$, $c_3$ and $\tau$ all fitted. This form is in agreement with other observations in the literature of an $\exp(-S^2)$ tail in the avalanche size distribution~\cite{Zaiser2007, Talamali2011}. Figure~\ref{avalanches_steady_state} shows the fits. Fitting data for all $k$ values with a shared exponent $\tau$ gives $\tau=1.342\pm0.004$ for the image sum kernel. For the Fourier kernel, we obtain a value  $\tau=1.364\pm0.005$ for $k=0.1$. These values match the trend in the stress-controlled data and are intermediate to the mean field value of $1.5$ and the value of $1.25$ reported by Talamali {\it et al}~\cite{Talamali2011}. 

In order to shed some light on this apparent discrepancy with previous results, we have also tested our data fitting routines on avalanche distributions obtained for a system of size $L=128$ and governed by the image sum kernel. This system size is comparable to that studied by Talamali {\it et al} with $L=256$. We find that the smaller systems yield an exponent of $1.31\pm0.01$ when fitted in the same manner as our large simulations, a value which is consistent with Talamali {\it et al}'s  $1.25\pm0.05$. It is most likely that this apparent change in exponent with system size is an artefact of least squares fitting with a limited power law regime, rather than a physical result, because the fit of our $L=128$ avalanche size distribution is improved when we fix the exponent to its $L=1024$ value of $\tau=1.34$. 

\begin{figure}
\centering
\includegraphics[width=\columnwidth]{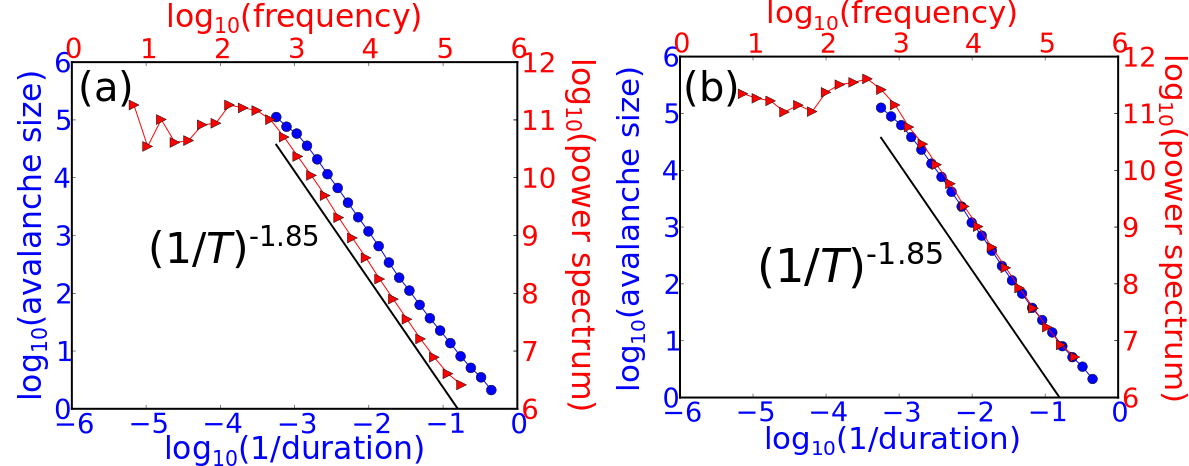} 
\caption{\label{temporal} Avalanche size {\it vs} inverse duration, and power spectrum of avalanche temporal signal, for the steady state of a system governed by (a) the image sum kernel and (b) the Fourier kernel, measured a strain-controlled simulation with $k=0.1$. Lines are guides to the eye, and the slope of $T^{-1.85}$ is an estimate based on visual comparison rather than a rigorous fit.}
\end{figure}

We also gather statistics on the temporal behavior of avalanches, which are shown for $k=0.1$ in Fig.~\ref{temporal}. The other $k$ values we study yield indistinguishable distributions. Like the size distributions, these are not consistent with mean field theory: the power spectrum of the avalanche signal and the mean avalanche size as a function of duration are given by a power law with measured exponent $1/\sigma\nu z \approx 1.85$ for both kernel implementations, compared to the mean field value of $2$~\cite{Dahmen2011}. The exact value of the exponent is uncertain because the power law scaling regime is rather narrow, less than 2 decades in $T$,  but the average avalanche size and the power spectrum appear to share the same exponent, as
expected theoretically \cite{kuntz2000}, and visual comparison indicates that the exponent is in the range $1.8 < 1/\sigma\nu z < 1.9$.  The distribution of avalanche durations also has a non mean-field exponent: fitting $P(T)$ with a power law $T^{-\alpha}$ gives $\alpha=1.5\pm0.09$ for the image sum kernel and $\alpha=1.5\pm0.09$ for the Fourier kernel, compared to the
 mean-field value $\alpha=2$ \citep{Dahmen2011}. The results for critical exponents are summarized in Table~\ref{exponents_comparison}.

\begin{table}
\centering
  \begin{tabular}{ | c | c | c | c | }
    \hline
Exponent & 2d amorphous & Mean field &  1d $1/r^2$ \\
	\hline
	$\tau$  & $1.342\pm0.004$ & $3/2$ & $1.25\pm0.05$ \cite{Bonamy2008a,Laurson2010} \\
	\hline
	$1/\sigma\nu z$  & $1.85\pm0.05$ & 2 & $\sim 1.7$ \cite{Bonamy2008a} \\
	\hline
	$\alpha$  & $1.5\pm0.09$ & 2 & $\sim 1.43$ \cite{Bonamy2008a} \\
	\hline
	$1/\sigma$  & $2.3\pm0.05$ & 2 & $2.1\pm0.08$ \cite{Laurson2010} \\
	\hline
  \end{tabular}
  \caption{Measured critical exponents, along with their values in the mean field and 1d $1/r^2$ depinning universality classes. Error bars are given where known quantitatively, except for the mean field results which are exact.}
\label{exponents_comparison}
\end{table}

\subsection{Strain localization near and far from criticality}

\begin{figure}
\centering
\includegraphics[width=0.95\columnwidth]{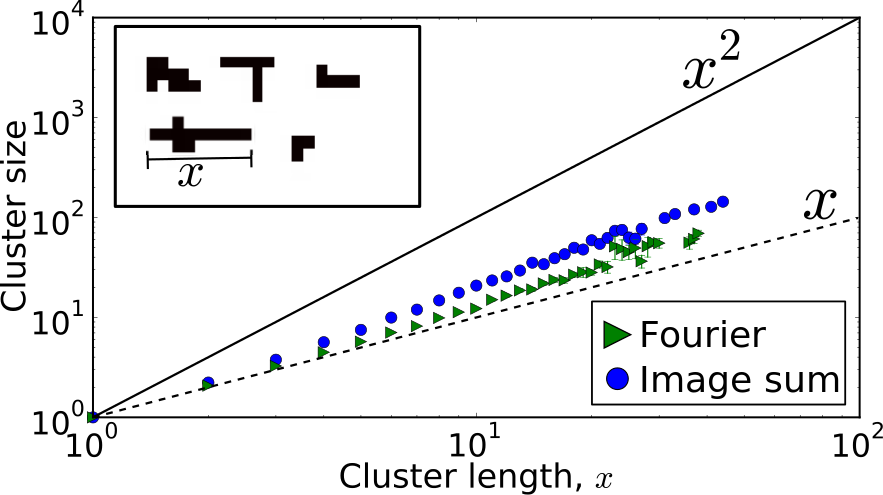} 
\caption{\label{clusters} Avalanche cluster size (number of sites) {\it vs} cluster length, for both kernels, measured at criticality, in strain controlled simulations with $k=0.1$. As indicated by the black lines, the scaling of cluster size with length is described by an exponent between $1$ and $2$. For both kernels, data from strain controlled simulations with $k=0.1$ has been used. The Inset shows typical non-rectangular clusters, and the definition of cluster length.}
\end{figure}

In addition to the spatially averaged avalanche activity discussed above, we have also analyzed the morphology of avalanches. Due to the long-range interactions, these may consist of several clusters, and we focus on the morphology of individual clusters. Typical examples of these are shown in Fig.~\ref{clusters}, along with the dependence of cluster size (given by the total number of sites) with cluster length, measured at criticality in strain controlled simulations. We find a power law relationship between, with an exponent slightly larger than $1$. We also find that the two kernel implementations lead to slightly different exponents; the Fourier kernel gives rise to avalanche clusters of slightly lower dimensionality. We note that in the definition used here, the size of a cluster is not equivalent to a strain, because a single site may participate in an avalanche more than once over the course of the avalanche duration. In our cluster size measurement, such a site is counted once, whereas its contribution to total strain is larger.

\begin{figure}
\centering
\includegraphics[width=0.95\columnwidth]{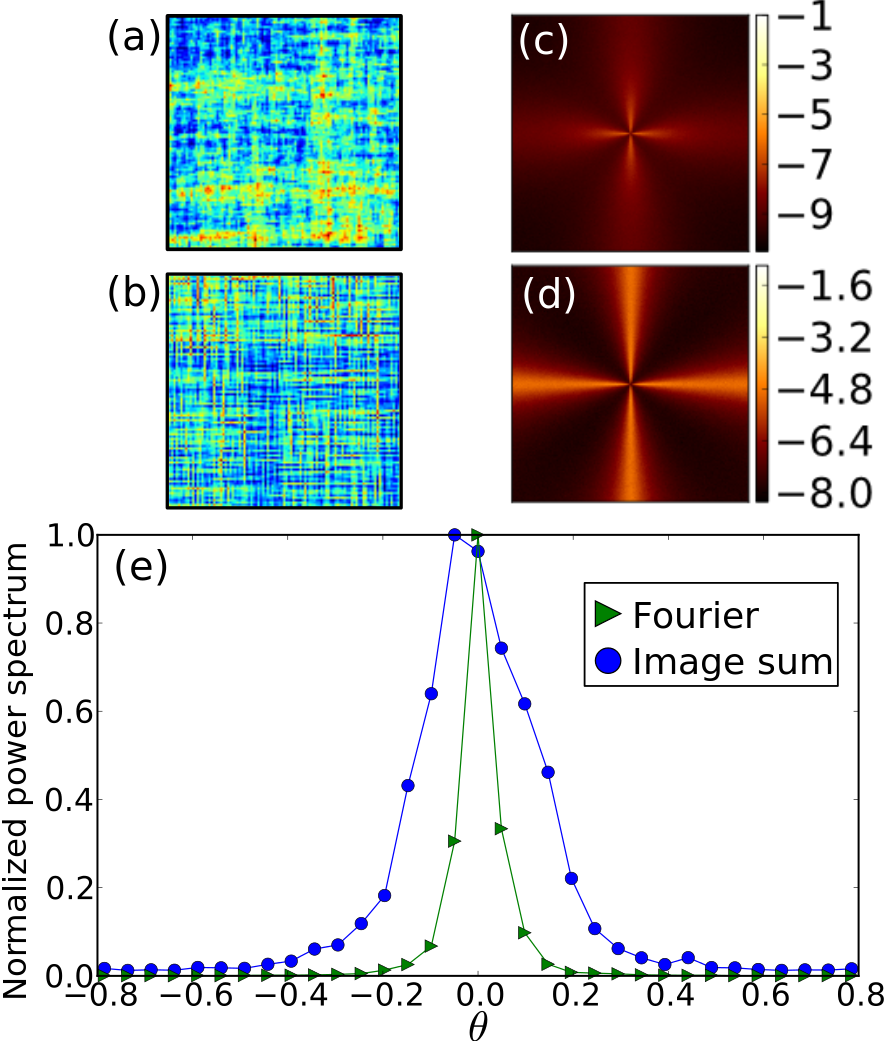} 
\caption{\label{morphology} $100\times100$ pixel samples of the strain distributions at criticality, for the image sum kernel (a) and Fourier discretized kernel (b). The localization is more pronounced for the Fourier kernel, as also seen in the power spectra of the strain maps obtained at criticality, averaged over $\sim 20$ realizations. The log (base 10) of these are shown for both the image sum kernel (c) and the Fourier kernel (d). The angular dependence, measured at $q=100$ is also shown in (e). The curves have been normalized by their peak value. For both kernels, data from strain controlled simulations with $k=0.1$ has been used.}
\end{figure}

\begin{figure}
\centering
\includegraphics[width=0.9\columnwidth]{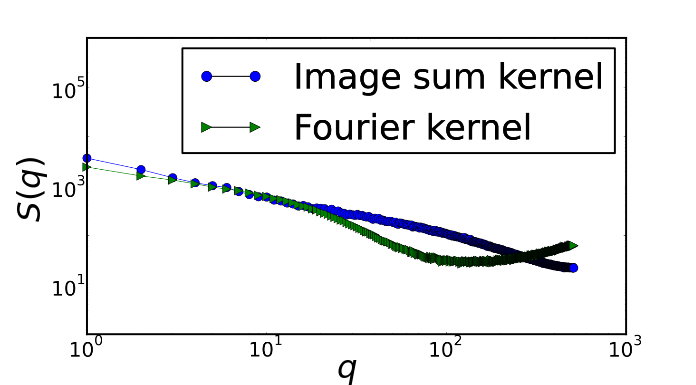} 
\caption{\label{power_spectrum_slice} Mean power spectra of horizontal and vertical 1d slices, taken at criticality. The ``drop off'' in $S(q)$ --- occurring at $q\sim20$ for the Fourier discretized kernel and $q\sim100$ for the image sum kernel --- indicates a length scale of the spatial structure of strain along a 1d slice. For both kernels, data from strain controlled simulations with $k=0.1$ has been used.}
\end{figure}

The effects of short-range interactions can be more clearly seen in the morphology of accumulated strain. Although both kernel implementations result in plastic activity localized into quasi-1d structures, as shown in Fig.~\ref{morphology} (a) and (b), the extent of this localization has a distinct dependence on short-range interactions. As previously noted by Talamali {\it et al}~\cite{Talamali2012}, the angular dependence of the power spectrum of the strain distribution displays a peak along the direction of localization. As shown in Fig.~\ref{morphology}(c)--(e), the peak is narrower for the Fourier discretized kernel than the image sum kernel. Indeed, as can be seen in Fig.~\ref{morphology}(a) and (b), the strain features generated by the Fourier kernel have a width of only a single lattice site, whereas the image sum kernel leads to strain morphology that does not appear to be so closely connected to the lattice. 

 Additional differences between the two kernels can be seen in the power spectra of 1d slices of the strain distribution. As shown in Fig.~\ref{power_spectrum_slice}, we find systems governed by the Fourier kernel have a longer correlation length --- the Fourier kernel has a scale of $q\sim20$, whereas the image sum kernel has a scale of $q\sim100$.

\begin{figure}
\centering
\includegraphics[width=0.95\columnwidth]{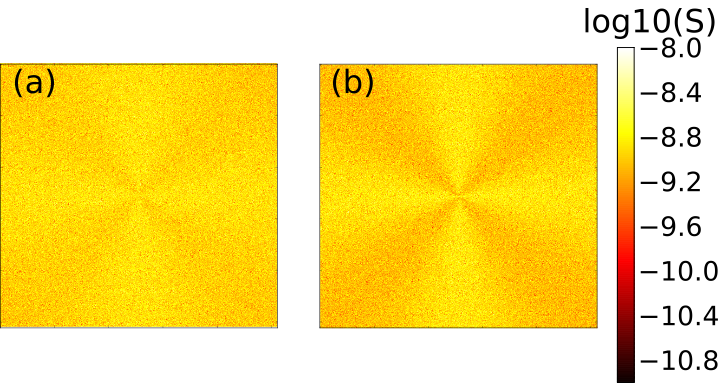} 
\caption{\label{small_strain_interface} Mean power spectra of strain distributions for systems governed by (a) the image sum kernel and (b) the Fourier discretized kernel, obtained at a mean strain of $0.05$, far from criticality. Although the data are too noisy for quantitative comparison, the spatial structure is visibly more pronounced for the Fourier kernel compared to the image sum kernel.}
\end{figure}

In fact, even at small stresses, the two kernels give rise to spatial distributions of strain with differing degrees of localization. For example, as shown in Fig.~\ref{small_strain_interface} when the mean strain in the system is $0.05$ --- far from criticality --- the power spectrum of the strain distribution of systems governed by the Fourier discretized kernel has visibly more spatial structure than that of the image sum kernel.

\subsection{Persistence of localized strain features}

\begin{figure}
\centering
\includegraphics[width=0.95\columnwidth]{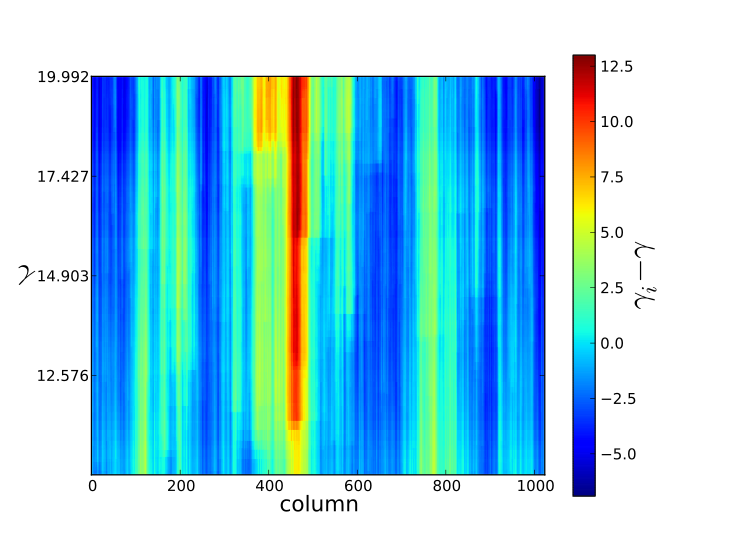} 
\caption{\label{fig5} Evolution of strain features in the critical steady state of a system governed by the image sum kernel, in a strain controlled simulation with $k=0.1$. ``Time'' (mean strain $\gamma$) increases moving upwards in each plot, while the horizontal axis represents space, in this case, columns of the $1024\times1024$ lattice. The color scheme indicates the difference between the mean strain of each column of the system, $\gamma_i$, and the mean strain of the whole system, $\gamma$. The data come from a typical single simulation run. Strains are given in units of the strain increment $d \gamma=0.1$.}
\end{figure}

It has previously been noted by Vandembroucq and Roux~\cite{Vandembroucq2011}, in their studies of a Fourier kernel, that features in the spatial distribution of strain are not permanent, but evolve as total strain increases. This is illustrated for a system governed by the image sum kernel in Fig.~\ref{fig5}, which shows the evolution of the difference between the mean strain of 1d slices of the system and the global mean strain. Features typically have a width of a few rows/columns. They can both appear and disappear, although appearance is more frequently observed than disappearance, and regions of particularly high strain (relative to the mean strain) can be long-lived. This is further illustrated in the movie of strain evolution given in the Supplemental Material~\cite{SupplementalMaterial}.

\begin{figure}
\centering
\includegraphics[width=0.95\columnwidth]{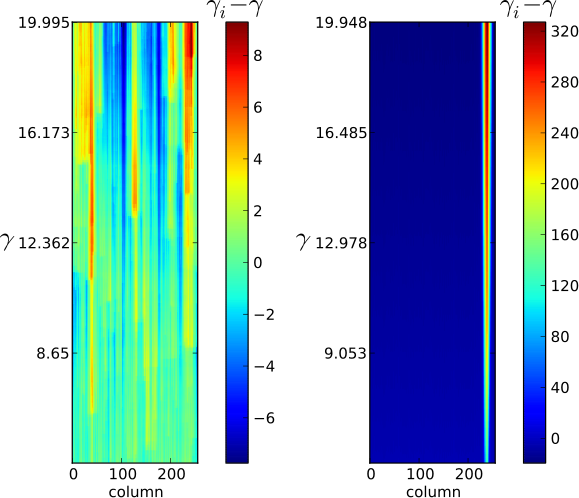} 
\caption{\label{weakening} Evolution of strain features in the critical steady state of systems governed by (a) the image sum kernel and (b) the Fourier kernel, in strain controlled simulations with $k=1$, in systems of size $L=256$. Both systems are subject to weakening as described in the main text, with $w=0.5$. ``Time'' (mean strain $\gamma$) increases moving upwards in each plot, while the horizontal axis represents space, in this case, columns of the $256\times256$ lattice. The color scheme indicates the difference between the mean strain of each column of the system, $\gamma_i$, and the mean strain of the whole system, $\gamma$. The data come from a typical single simulation run. Strains are given in units of the strain increment $d \gamma=0.1$.}
\end{figure}

Although both kernels give qualitatively similar results in the absence of weakening or hardening, when weakening is introduced the effect of short-range interactions becomes important. Like Vandembroucq and Roux~\cite{Vandembroucq2011}, we impose weakening by biasing the initial yield stress distributions by an amount $w$. In other words, the local yield stresses are initially drawn from a uniform distribution over $[w, w+1)$, but after yield events the renewed yield stresses are drawn from a uniform distribution over $[0, 1)$. As shown in Fig.~\ref{weakening}, the system governed by the Fourier kernel displays strong localization into a narrow, persistent ``shear band'', but the image sum kernel does not lead to such localization. The difference in strain between the ``shear band'' and the mean strain is an order of magnitude larger than the differences seen in other simulations, an indication that the feature is a lattice instability generated by the Fourier implementation of the kernel. In the Supplemental Material~\cite{SupplementalMaterial}, we include movies of the evolution illustrated in Fig.~\ref{weakening}.

\section{Conclusion}
To summarize: we have seen that plasticity in 2d amorphous materials has non mean-field critical behavior, which is universal. We have also seen that, depending on details of the interaction kernel, mean field behavior can be recovered away from the critical point. These two observations are linked by the localization of strain, which depends strongly on short-range interactions. While both kernels lead to localization into narrow features, the strain localization for the Fourier kernel appears to be closely related to the square lattice on which the model is discretized, with features having a width of a single lattice site. When strain is not localized --- as in the case of the image sum kernel away from criticality --- behavior consistent with mean-field depinning is recovered. 

Furthermore, as indicated by the striking difference between the two kernels in the weakening tests, nonuniversal features relating to strain localization are not simply curiosities, but also have implications for how the ``depinning'' model for amorphous plasticity can be applied to real materials. While long-lived, spatially-localized strain features are universal, the fact that their stability depends strongly on short-range interactions raises serious questions about the physical interpretation of ``shear bands'' in systems governed by the Fourier kernel. As noted already (see Fig.~\ref{kernels}), the Fourier kernel noticeably deviates from $1/r^2$ scaling in the first and second neighbor interactions on a square lattice, and behavior that derives from this deviation should be considered a spurious lattice effect.

Because of strain localization, one hypothesis for the non mean field behavior of the model is dimensional reduction, that is, as criticality is approached, the system becomes effectively a set of (almost) noninteracting 1D subsystems. However, the exponents $\tau\approx1.35$ and $1/\sigma \nu z \approx 1.85$ are not consistent with the values expected in a 1D system with $1/r^2$ interactions (i.e  $\tau=1.25$ and $1/\sigma \nu z=1.7$~\cite{Bonamy2008a,Laurson2010}; see Table~\ref{exponents_comparison}). Furthermore, as discussed above, avalanches at criticality have dimensionality above $1$.  In other words, although there is spatial localization of strain, we do not observe complete dimensional reduction, for either kernel. 
Intriguingly, recent simulation studies~\cite{Laurson2013} of magnetic domains in thin films with long-range
dipolar interactions report an avalanche exponent $\tau\simeq 1.33$. It remains an open question whether this model shares the universality class of two dimensional amorphous plasticity.

\begin{acknowledgments}
This work is supported by the European Research Council through the Advanced Grant 2011 SIZEFEECTS. We thank M. J. Alava, Z. Bertalan, G. Durin, L. Laurson and M. Zaiser for helpful discussions.
\end{acknowledgments}


\appendix

\section{Image sum kernel}
\label{ImageSum}
In Cartesian coordinates, the interaction kernel for an infinite system is
\begin{equation}
K(x, y) = \frac{x^4 + y^4 - 6 x^2 y^2}{(x^2+y^2)^3}.
\end{equation}
In order to periodize the kernel, we sum over images of period $L$, by making the substitution $x \to x + k_x L$, $y \to y + k_y L$ and summing over the $k$s. This sum is only conditionally convergent, and different orders of summation yield different results. However, we are primarily interested in generating an interaction kernel that is periodic, with period $L$, and that behaves like $\cos(4\theta)/r^2$ as much as possible. In other words, the summation is purely formal, and we can select an order of summation based on convenience.

We perform the periodization by summing over an infinite number of images in the $y$ direction, to give
\begin{widetext}
\begin{equation}
\begin{split}
K(x,y) = \sum_{k_x} \frac{\pi^2}{2 L^3} \biggl[ \bigl(2\pi (x+k_x L){\rm coth}(\frac{\pi}{L}(x+k_xL-i y))-L\bigr){\rm csch}^2(\frac{\pi}{L}(x+k_xL-i y)) \\
+ \bigl(2\pi (x+k_x L){\rm coth}(\frac{\pi}{L}(x+k_xL+i y))-L){\rm csch}^2(\frac{\pi}{L}(x+k_xL+i y)\bigr) \biggr].
\end{split}
\end{equation}
\end{widetext}
The terms in the summand are exponentially decaying in $|k_x|$, so we can approximate the sum by using a small number of terms (typically $-5 \leq k_x \leq 5$). Further manipulation of the hyperbolic functions yields an expression involving multiplication and addition of terms that depend only on $x + k_x L$ or $y$ individually, which allows efficient storage of the interaction kernel in memory (scaling as $L$ rather than $L^2$). This is important for implementation on a graphics processing unit, which has limited memory, and allows us to simulate systems of size $L=1024$ efficiently.

\section{Parallelization of the simulation algorithm}
\label{gpu}
For small systems, our algorithm (described in Section~\ref{methods}) can be implemented on a CPU, but because every site interacts with every other site, the computational time to obtain a target strain scales as $L^4$.

There are three points at which we parallelize this algorithm. First, the detection of the site closest to its threshold can be trivially parallelized, as it is a reduction problem. We use the function provided in the {\tt CUBLAS} library. Second, updating the strain and the yield threshold of sites during an avalanche is also performed in parallel. We use the CUDA random number generator {\tt curand}, and have verified that the thresholds generated are ``well behaved'', with no correlations appearing in the sequence of thresholds at each site.

The key point of parallelization, however, is in the updating of local stresses due to interactions. This is also the most delicate point, as care must be taken to avoid race conditions. We allocate a separate thread to each site that has not increased its strain, and calculate (from lookup tables) the effect on the total stress at each site from the sites that have increased their strain. The corresponding ``restoring force'' on the sites that have updated can be calculated at the same time, and is added using CUDA's built-in {\tt atomicAdd}, since every thread has a contribution to its value.
\end{document}